\documentclass[printer]{aa}

\usepackage{graphicx}

\begin{document}

\def\P{\bar{\Phi}}

\def\st{\sigma_{\rm T}}

\def\vk{v_{\rm K}}

\def\sles{\lower2pt\hbox{$\buildrel {\scriptstyle <}
   \over {\scriptstyle\sim}$}}

\def\sgreat{\lower2pt\hbox{$\buildrel {\scriptstyle >}
   \over {\scriptstyle\sim}$}}

\title{The magnetic field topology in the reconnecting
pulsar magnetosphere}

\author{Ioannis Contopoulos}
\institute{Research Center for Astronomy, Academy of Athens, 
GR-11527 Athens, Greece, 
\email{icontop@academyofathens.gr}}

\titlerunning{Magnetic topology}

\date{Received January 2007 / Accepted April 2007}

\abstract{
We show that toroidal magnetic field
annihilation in the equatorial current sheet
of the pulsar magnetosphere is related to 
how fast poloidal magnetic field lines close as we move
away from the light cylinder.
This determines the radial reconnection electric field
which directly accelerates particles in the neutral
layer inside the equatorial current sheet.
The efficiency of the poloidal magnetic field closure
near the light cylinder
may be measurable through the pulsar braking index.
We argue that the lower the efficiency of
pair formation, the higher the braking index.
We also argue that the synchrotron radiation reaction
in the neutral layer does not inhibit the
accelerated particles from reaching the maximum
energy of about $\sim 10^{16}$~eV that is
available in the open pulsar magnetosphere.

}

\maketitle

\section{Introduction}

After almost 40 years of pulsar research, the structure
of the pulsar magnetosphere, the region extending from
the neutron star surface to the wind termination shock,
is still not clearly understood. 
The observed pulsed radiation (radio to gamma ray)
is believed to originate in the magnetosphere, and yet it
is still not clear where and how it is produced
(e.g. \cite{A83}; \cite{MH04}).
The magnetosphere is also
the source of a fast particle wind with
Lorentz factor on the order of $10^6$ that
terminates in a standing shock at a distance of about 
0.1~pc (\cite{BA00}).
Several acceleration mechanisms have been proposed, ranging
from ideal MHD models (\cite{CK02}; \cite{V04}), pressure driving 
(\cite{C90}; \cite{M94}; \cite{KS03}), 
to direct particle acceleration in
the equatorial current sheet (\cite{RCL05}; \cite{C07}), 
but we still do not understand how this wind is produced. 

In recent years, there has been
mounting evidence against ideal MHD wind acceleration,
and there is no convincing self-consistent, ideal MHD 
wind solution available in the literature (\cite{B97}).
Thus, we decided to direct our attention to 
models that contain regions where ideal MHD breaks down.
The most promising of these regions is the equatorial current sheet,
across which the main
magnetic field component, the toroidal one, changes direction
and annihilates (\cite{C07}). 
Annihilation of the toroidal magnetic field
component is associated with an electric field that
accelerates particles directly inside the current sheet.
Although several authors have acknowledged the presence of this
electric field (e.g. \cite{K04}; \cite{RCL05}), 
we feel that a certain
important physical element has been neglected in the discussion, 
namely the global topology of the poloidal magnetic field.

In \cite{C07}, we dealt with non-ideal MHD regions through
the notion of the effective magnetic diffusivity.
In the present work, we give a simpler picture without
mentioning the magnetic diffusivity in the equatorial
current sheet.
In \S~2 we argue that toroidal magnetic field
annihilation cannot be studied independently
of the poloidal magnetic field. In \S~3 we associate the
pulsar braking index to the rate at which poloidal magnetic-field 
lines close with distance in the vicinity of the light cylinder.
Finally, in \S~4 we discuss the issue of
particle acceleration in the equatorial current sheet.

\section{The global topology of the pulsar magnetosphere}

Let us introduce a spherical system of coordinates $(r,\theta,\phi)$
centered on the neutron star and aligned with its axis of rotation.
The neutron star rotation delineates two distinct
hemispheres, the upper and lower, separated by the rotation equator.
The magnetic field also delineates two half spaces: the north,
where the magnetic field points away from the star, 
and the south, where the magnetic field points towards the star.
In general, the rotation and magnetic axes are at an
angle $\vartheta \neq 0$, and we may
assume without any loss of generality that $\vartheta < 90^o$.
In that case, $B_r$ is mostly positive in the upper (`north')
hemisphere, whereas it is mostly negative in the lower (`south')
hemisphere.

In the case of an isolated
pulsar, all magnetospheric field lines naturally originate
on the surface of the neutron star, since the source
of the magnetospheric field is the neutron star interior. 
Obviously, solutions also exist with field lines not connected 
to the star (\cite{LTR06}), and these may very well apply to 
neutron stars surrounded by disks with disk winds,
but not to isolated neutron stars.
The star holds the magnetic field from diffusing to infinity.
The field close to the neutron star
is dipolar, but as we move away from the stellar surface, 
the magnetic field topology is gradually distorted.
The magnetic field is
stretched outwards and is wound backwards because of the neutron
star's rotation. The distortion depends on the magnetospheric
plasma conductivity, on the inclination angle between
the rotation and magnetic axes, and, as we will see, on the
neutron star spindown history. 
The above characteristics can be found in several
numerical solutions of simple limiting cases (e.g. \cite{CKF99};
\cite{S06}).

Let us now discuss the significance of the magnetospheric plasma
conductivity, 
first considering the case of infinite plasma conductivity.
Ideal MHD requires that plasma is confined to flowing along
the magnetic field, so field lines may be visualized
as corrotating with the neutron star at the stellar
angular velocity $\Omega$. This does not mean that the 
magnetospheric plasma corrotates with the star.
The plasma angular velocity is equal to $\Omega$ only
wherever $B_\phi=0$ (e.g. \cite{Be97}). A direct corollary of this is that
no field lines close outside the light cylinder, which is
the cylindrical distance $r_L\equiv c/\Omega$, where if a particle
were to corrotate with the neutron star, it would
do so at the speed of light. 
The argument is straightforward: as we said,
field lines from both stellar magnetic hemispheres are wound
backwards by the neutron star's rotation; therefore, a point
on every closed field line may be found where $B_\phi=0$. 
If a closed field line were to extend beyond the light cylinder,
that point would lie outside the light cylinder, and
according to our discussion above, the plasma velocity there would
have to be at least equal to $r\Omega \sin\theta >c$.
The unavoidable conclusion is that 
the pulsar magnetosphere consists of closed field lines that
corrotate with the neutron star inside the light cylinder
and of field lines that cross the
light cylinder and open up to infinity.
It is important to keep in mind that nothing
exceptional takes place at the light cylinder, because
field lines cross it without kinks or discontinuities 
(\cite{CKF99}).
Several authors have shown that asymptotically ($r\sgreat r_L$), 
the poloidal magnetic field structure
becomes radial (monopole-like). 
In the axisymmetric case ($\vartheta\approx 0$),
the open poloidal magnetic flux in each
magnetic hemisphere is given by
\begin{equation}
\Psi(r,\theta)\approx \Psi_L(1-\cos\theta)\ ,
\label{Psi}
\end{equation}
and the magnetic field by
\[
|B_r (r,\theta)|\approx B_L \left(\frac{r}{r_L}\right)^{-2}\ ,
\ B_\theta (r,\theta)\approx 0\ \mbox{and}\ 
\]
\begin{equation}
|B_\phi (r,\theta)|\approx B_L \sin\theta
\left(\frac{r}{r_L}\right)^{-1}
\label{B}
\end{equation}
(\cite{M94}; \cite{CKF99}). Here,
$B_L\equiv \Psi_L/(2\pi r_L^2)$ is a typical value for the magnetic field
at about the light cylinder distance.
As \cite{B99} shows in a very important and elegant
paper, Eqs.~\ref{B} are valid even in the case
of an oblique (i.e. non-aligned) magnetic rotator.
In that case, in the upper
hemisphere, magnetic field lines mostly point
away from the star ($B_r>0$, $B_\phi<0$),
whereas in the lower hemisphere they
mostly point towards it ($B_r<0$, $B_\phi>0$). 
The two regions are separated by a spiral current-sheet 
discontinuity described by Eq.~3 in Kirk \&
Lyubarski~(2001), 
across which $B_r$ and $B_\phi$ change sign together.

Let us next consider the more realistic case of high, but not
infinite, plasma conductivity. 
It is natural to argue that
ideal MHD is a realistic description of the largest part of
the pulsar magnetosphere, except for restricted regions
with high electric-current densities, where ideal MHD
breaks down (\cite{C07}). In that case, field lines are
not forbidden to close outside the light cylinder.
In fact, several numerical solutions 
that try to simulate the ideal MHD magnetosphere
show this different topology because of numerical diffusivity
(\cite{K06}; \cite{McK06}; \cite{S06}). 
It is very important to note that
field-line closure outside the light cylinder may take
place only if it is associated with magnetic field energy
dissipation (non-ideal MHD). In fact, several authors have
proposed that this mechanism may be related to the
acceleration of the pulsar wind, and this led to the
so-called `striped' pulsar wind model
(\cite{C90}; \cite{M94}; \cite{LK01}; \cite{KS03}).

There is here, we believe, a subtle but very important point
that has been overlooked by most previous studies of
magnetic field dissipation in the current sheet outside the
light cylinder, and this has to do with the global topology
of the reconnecting magnetic field. Equations~\ref{B}
imply that, next to the region of
reconnection in the striped wind at large
distances ($r\gg r_L$), it is the toroidal magnetic 
field component that is the dominant one. Several authors,
therefore, opted to ignore the poloidal magnetic field component
altogether and instead study reconnection as if it were
taking place exclusively in the toroidal magnetic field component.
It is as if 
the star `shoots out' rings of a toroidal magnetic field
with different $\phi$-direction depending on whether they
originate in the north or south magnetic hemispheres.
The problem with that approach 
is that, by ignoring the global topology of the poloidal magnetic field
component, we miss the crucial fact that the source of the
toroidal magnetic field component is the azimuthal winding
of the poloidal one.

This is not without significance.
All field lines originate on the surface of the star, and
there is a finite amount of magnetic flux that crosses
the light cylinder and would be stretched out to infinity,
were it not for the presence of equatorial current sheet
reconnection.
Obviously, as toroidal magnetic field annihilation
proceeds, less and less poloidal
magnetic flux extends to larger distances\footnote{
In other words, toroidal magnetic field annihilation
proceeds in the north-south direction.}. As a result, less and
less open field lines are available for reconnection or for
carrying with them the stellar spindown energy in the
form of Poynting flux. We may thus generalize Eqs.~\ref{Psi}
\& \ref{B} and introduce a simple power law scaling
\begin{equation}
\Psi(r,\theta)\approx \Psi_L (1-\cos\theta)
\left(\frac{r}{r_L}\right)^{-\epsilon}\ ,
\label{Psi2}
\end{equation}
where $\epsilon\ll 1$ is a small positive parameter
that obviously depends on the (yet unknown) 
physics of gradual reconnection
in the equatorial sheet. At this stage, we can only take it
to be a free parameter. Equation~\ref{Psi2} is
an ad hoc generalization of Eq.~\ref{Psi}, and is 
not based on current theory
or observations. It may, however, be valid as a toy model
that describes toroidal magnetic field annihilation
and poloidal magnetic field reconnection in the
vicinity of the light cylinder. Obviously, in order to study
 how fast toroidal magnetic field
annihilation proceeds with distance, one needs to solve
a formidable non-force-free (beyond the fast magnetosonic surface)
relativistic MHD problem that incorporates the
microphysics of field annihilation. It is
interesting to note that recent time-dependent
numerical simulations (\cite{BA06}) suggest that
reconnection may be sporadic, spontaneous, and limited
in the vicinity of the closed line zone.

Having acknowledged the limitations of our toy model,
it is now straightforward to show that
\[
|B_r (r,\theta)|\approx B_L \left(\frac{r}{r_L}\right)^{-2-\epsilon}\ ,\ 
B_\theta (r,\theta)\approx \epsilon B_L 
\left(\frac{r}{r_L}\right)^{-2-\epsilon}\ \mbox{and}\ 
\]
\begin{equation}
|B_\phi (r,\theta)| \sim B_L \sin\theta
\left(\frac{r}{r_L}\right)^{-1-\epsilon}\ .
\label{B2}
\end{equation}
Note that $B_\theta$ is everywhere of the same sign. 
This component of the magnetic field,
which up to now has been considered insignificant,
may have interesting observable implications.
In fact, \cite{PK05} present a plausible physical model
for the Crab pulsar, where high energy radiation
originates in the equatorial current sheet
at a distance of about $10 r_L$. In their model,
they require a magnetic field component $B_\theta$
roughly 10\% of $B_\phi$, which is not very
different from what Eqs.~\ref{B2} yield for 
$\epsilon\sles 1$.

We conclude this section by emphasizing that annihilation
of the toroidal magnetic field in the equatorial current sheet
may be studied
only in conjunction with studying the poloidal structure
of the magnetospheric field. To our knowledge, this has
not been taken into account in any existing model of
extended equatorial field annihilation (e.g. \cite{LK01}; 
\cite{RCL05}).

\section{Evolution of the pulsar magnetosphere
with neutron star spindown}

As we argued in Contopoulos~(2007), the rate of equatorial
current-sheet reconnection in the vicinity 
of the light cylinder may be indirectly measurable
through observations of the neutron star's braking index 
\begin{equation}
n\equiv \frac{\delta\ln \dot{\Omega}}{\delta \ln\Omega}\ .
\label{n}
\end{equation}
In that work, we dealt with non-ideal MHD regions through
the notion of the effective magnetic diffusivity.
In the present work, we give a simpler picture without
mentioning the magnetic diffusivity in the equatorial
current sheet.

In the above picture of finite plasma conductivity,
magnetic field lines
enter the equatorial current sheet at a grazing
angle, i.e. $|B_r|\gg B_\theta$ outside
the current sheet. Electromagnetic energy
is continuously `pumped' into it and is lost in
accelerating particles in the current sheet (see next section). 
In other words, equatorial toroidal magnetic field annihilation
prevents poloidal magnetic flux from reaching `infinite' distances.
The picture we describe here may even be thought to take
place in steady state were it not for the fact that
the neutron star actually spins down.

Why is the neutron star spindown so important?
We remind the reader that, inside the light cylinder,
the pulsar magnetosphere consists of a corrotating region
of closed field lines and of a certain amount $\Psi_{\rm open}$
of open poloidal magnetic flux
which extends beyond the light cylinder.
It is well known that the spindown rate $\dot{\Omega}$ depends
on $\Omega$ and $\Psi_{\rm open}$ as
\begin{equation}
\dot{\Omega} \propto \Omega \Psi_{\rm open}^2
\label{Omegadot}
\end{equation}
(\cite{C05}).
Let us backtrack for a moment and assume that reconnection
does not take place anywhere in the pulsar magnetosphere.
In that case, as the neutron star spins down, the light
cylinder moves out at a certain rate (roughly 1~km
per year), but the region of closed field lines
cannot grow, since the absence of reconnection
prevents open field lines from closing and
accumulating on top of the closed line region,
and thus Eq.~\ref{Psi} implies that
$\Psi_{\rm open}\equiv \Psi_L=\mbox{const}$. 
According to Eqs.~\ref{n} \&
\ref{Omegadot}, the neutron star braking index would be
measured to be equal to 1, which is not what is observed.
Now let us once again consider the physically
more interesting case where reconnection does take
place in the vicinity of the light cylinder, only it is
confined inside the equatorial current sheet.
In that case, as the neutron star spins down
and the light cylinder moves out, it encompasses
more and more poloidal magnetic flux because, as
we argued, in the vicinity of the light cylinder
magnetic flux closes with distance at a rate
given by Eq.~\ref{Psi2}.
When the light cylinder moves out from $r_L$
to $r_L+\delta r_L$, a certain amount of poloidal magnetic flux,
\begin{equation}
\delta \Psi_{\rm open}\approx -\epsilon \Psi_L
\left(\frac{\delta r_L}{r_L}\right)
= \epsilon \Psi_L
\left(\frac{\delta \Omega}{\Omega}\right)\ ,
\label{rate}
\end{equation}
is now found contained inside the new light cylinder.
Under ideal MHD conditions, however, the stellar 
spindown energy is transferred only along field lines that 
cross the light cylinder\footnote{
As we show in the next section, 
a radial electric field equal to 
$E_r=B_\theta (r/r_L)$ appears inside the equatorial
current sheet. Obviously, inside
the light cylinder, ${\bf E}\perp {\rm B}$ and 
$E_r<B$, so electromagnetic field energy cannot be
dissipated through direct particle acceleration 
in the equatorial current sheet.
}; therefore, our result in Eq.~\ref{rate}
modifies the neutron star spindown rate by an amount
\begin{equation}
\delta \dot{\Omega}=(1+2\epsilon)\frac{\dot{\Omega}}{\Omega}
\delta\Omega\ ,
\end{equation}
which yields a braking index value of
\begin{equation}
n=1+2\epsilon\ .
\end{equation}

The reader may have noticed that our present discussion
differs from the one in Contopoulos~(2007), where 
we introduced an effective magnetic diffusivity $\eta$, which is
responsible for both the equatorial reconnection of the toroidal
magnetic field and the inward diffusion of the poloidal
magnetic field lines.
In the present work, we again consider 
magnetic field energy losses in the equatorial
current sheet, but we also argue
that the growth of the closed-line region is directly
related to the advancement of the light cylinder to
larger distances as the neutron star spins down.

We repeat once again that toroidal magnetic field annihilation
is related to how fast poloidal magnetic field lines reconnect 
as we move away from the light cylinder (Eqs.~\ref{Psi2}
\& \ref{B2}). This is very important, because
the efficiency of poloidal magnetic field closure is
something that may be measurable through the pulsar
braking index, and thus yield an estimate
of the physical processes that lead to
magnetic field reconnection in the pulsar
magnetosphere.

\section{Particle acceleration in the equatorial current sheet}

Equations~\ref{B2} are valid outside the current sheet discontinuity. 
Inside the sheet, $B_r$ and $B_\phi$ change sign, meaning
a neutral layer develops where the only remaining magnetic field 
component is $B_\theta$. 
In other words, when we generalize
Bogovalov~(1999), reconnection may take place at various
positions along the current sheet discontinuity.
All field lines originate on the
surface of the star; therefore, as we argued, they carry
with them the information that the neutron star is rotating
at angular velocity $\Omega$ throughout the region where
ideal MHD is a valid physical approximation.
In axisymmetry, a poloidal electric field develops perpendicular
to the poloidal magnetic field because of the neutron
star rotation, with magnitude
\begin{equation}
E=\frac{r\Omega\sin\theta}{c}\left(B_r^2+B_\theta^2\right)^{1/2}
\label{E}
\end{equation}
(e.g. \cite{Be97}). Magnetic flux surfaces are, therefore,
isopotential surfaces, with potential difference between them
given by
\begin{equation}
\delta V=\frac{\Omega}{2\pi c}\delta \Psi\ ,
\end{equation}
where $\delta\Psi$ is the magnetic flux contained between
the two flux surfaces. This is valid along
each field line, all the way from its footpoint on the stellar
surface to the point where that field line reaches the 
equatorial current sheet and reconnects 
(if that particular one does) with the corresponding
field line from the opposite magnetic hemisphere.
According to our discussion above, this implies that
the magnetospheric potential drop manifests itself along the
equatorial sheet wherever poloidal field lines reconnect.
Another way to understand this is that,
wherever reconnection takes place, a radial electric field equal to
\begin{equation}
E_r=\frac{r\sin\theta}{r_L}B_\theta
\label{Er}
\end{equation}
develops in the
interior of the current sheet where $B_r\sim B_\phi \sim 0$.
This is the reconnection electric field component mentioned in
Kirk~(2004) and Romanova, Chulsky \& Lovelace~(2005)
that accelerates particles in the current sheet. 
The new result in our present work is that $E_r$
is directly proportional to $B_\theta\neq 0$ 
in the current sheet.

As long as a particle
stays in the interior of a reconnection region in
the equatorial current sheet,
it will be accelerated to an energy equal to
$e\delta V$, where
$e$ is the electron charge, and $\delta V$ is the
potential drop associated with the poloidal field lines at the
two ends of the acceleration region.
Current sheets are known to be unstable, 
so we expect that the pulsar equatorial
current sheet will also
be fragmented into several short acceleration regions.
Nevertheless, one may argue that
some particles may manage to profit from several
of these acceleration regions. In such a stochastic acceleration,
some particles may manage to get accelerated
to the maximum energy available in the open pulsar magnetosphere,
i.e. to the energy
due to the potential drop across field lines that originate
on the magnetic pole and the edge of the neutron star polar cap, 
namely $\sim 10^{16}\ \mbox{eV}$ (\cite{DeJH92}).

We argue that synchrotron radiation reaction does not
inhibit the acceleration. We performed a straightforward
calculation of relativistic electron/positron orbits 
as they enter the current sheet (Figs.~1 \& 2) and realized that their orbit
becomes more and more confined around the center of the
current sheet where $B_r\sim B_\phi \sim 0$. Therefore, 
even if one is willing to keep our toy model scaling
all the way from $r_L$ to $r\gg r_L$, 
the synchrotron radiation reaction 
due to the $B_\theta$ component of the magnetic field
will limit the acceleration only
when the particles reach a Lorentz factor on the order of
\begin{equation}
\Gamma_{\rm max}\sim 10^7\left(
\frac{r}{r_L}\right)^{1/2}B_{L}^{-1/2} \gg 10^7\ ,
\end{equation}
where, $B_L$ is measured in Gauss. Thus, synchrotron
radiation is unimportant inside the pulsar wind
termination shock.

We end our discussion with an order-of-magnitude 
estimate of the acceleration length $l_{\rm acc}$
in the vicinity of the light cylinder.
Electromagnetic energy flux enters the current sheet
from above and below at a rate of
$2(E_r B_\phi/4\pi c)(B_\theta /B_r)$.
This is absorbed by the particle flux 
$2\kappa n_{GJ} c (B_\theta /B_r)$, which also
enters the current sheet from above and below.
Here, $\kappa\sim 10^3-10^5$ is the multiplicity of pair
production in magnetospheric gaps (\cite{DH82}; \cite{GI85}), 
and $n_{GJ}\sim \Omega B_r/(4\pi ce)$ is the
Goldreich-Julian number density (\cite{GJ69}).
If we assume an acceleration length $l_{\rm acc}\ll r$, 
a straightforward energy balance in the current sheet yields
\begin{equation}
l_{\rm acc}\sim \frac{r}{2\kappa \epsilon}\ .
\label{lacc}
\end{equation}
A (typical) value of $\epsilon$ on the order of one half
yields $l_{\rm acc}\sim 10^{-4} r\ll r$.
On the other hand, if we know
$l_{\rm acc}$ from a numerical simulation of the
pulsar current sheet fragmentation 
(e.g. \cite{Cetal01}; \cite{BB05}), 
Eq.~\ref{lacc} yields $\epsilon$, which
determines the global topology of the pulsar magnetosphere
for various values of the multiplicity parameter $\kappa$.
We thus predict that the smaller the efficiency of pair production,
the larger $\epsilon$ is. Unfortunately,
the 6 pulsars with the best-measured values of the
braking index are all strong pair creators, and
therefore, our prediction cannot be tested observationally
using only those particular 6 pulsars.

\section{Summary}

In the present work, we argue that toroidal magnetic
field annihilation in the equatorial current sheet of the
pulsar magnetosphere cannot be studied independently of
the poloidal magnetic field. This association
allows us to relate the pulsar braking index to the
rate at which poloidal magnetic field lines
close with distance in the vicinity of the light cylinder.
This also allows us to estimate the reconnection
electric field that develops in the interior
of the equatorial current sheet.
We show that this electric field accelerates
electrons and positron in opposite directions, and 
we predict that there will be little or 
no synchrotron radiation in the region
upstream of the wind termination shock. In fact,
some particles may even profit from the full potential
drop in the open line region and thus reach energies
on the order of $10^{16}$~eV.
Finally, we associate the efficiency of equatorial
reconnection to the efficiency of pair production
in the pulsar magnetosphere.

\acknowledgements{We would like to thank Pr. Roger Blandford
for his hospitality at the Kavli Institute
for Particle Astrophysics and Cosmology
in July 2006 when most of the
ideas in the present work originated.}

\newpage

\begin{figure}
\includegraphics[angle=270,scale=0.5]{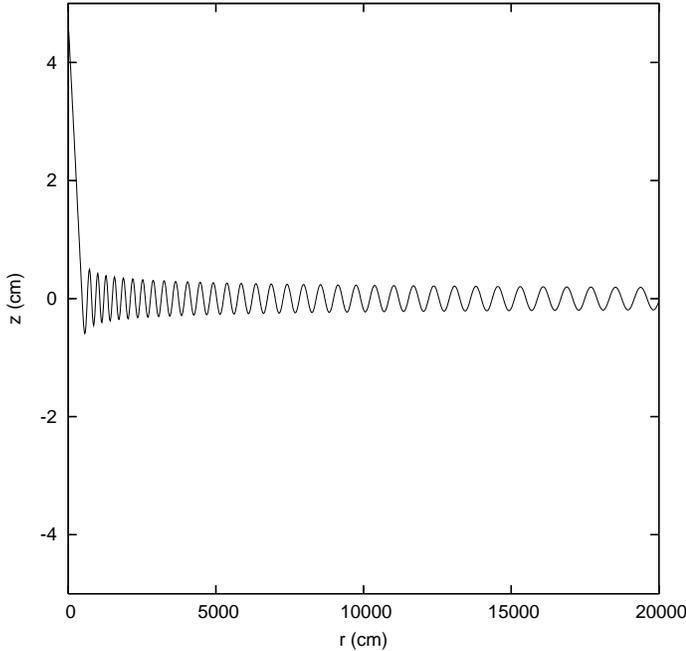}
\caption{Calculated
electron orbits that enter the equatorial current
sheet $z=0$ from above. The units are in centimeters. 
We consider only the axisymmetric case 
$\vartheta=0$. The injection region is at $r=10r_L$.
Here, $B_\theta/B_r=10^2$ on the surface of the current sheet.
We took $r_L=1500\mbox{km}$ and $B_L=10^6\mbox{G}$ (values
representative of the Crab pulsar).
Initially, the relativistic electrons move together
with the magnetic field with Lorentz factor equal to $\sim 100$.
The Lorentz factor grows linearly with the acceleration distance.
We observe that the particles do not escape from the current sheet
(unless of course the current sheet is terminated), and 
are confined more and more around the center of the
current sheet where ${\bf B}=B_\theta\hat{\theta}$.}
\label{fig1}
\end{figure}

\begin{figure}
\includegraphics[angle=270,scale=0.5]{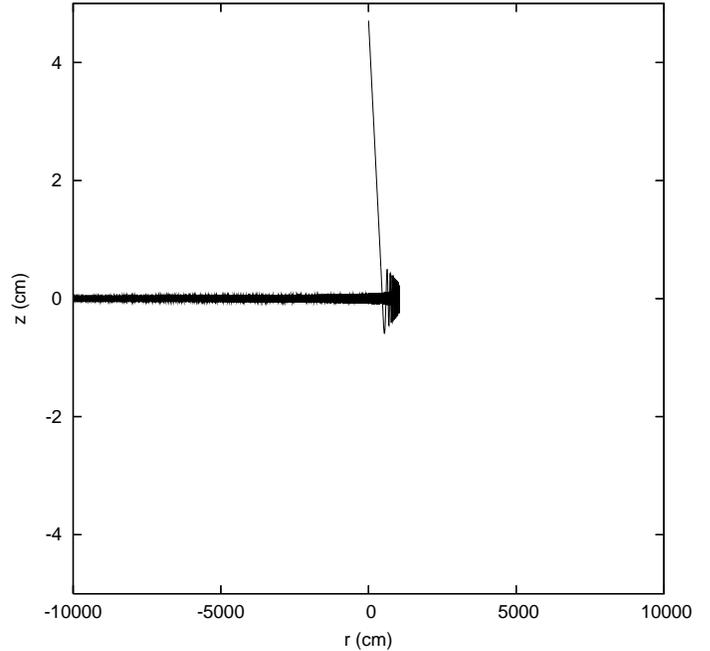}
\caption{Same as Fig.~1 for positrons.}
\label{fig2}
\end{figure}

\end{document}